\documentclass{elsart}
 \usepackage{graphics}
 \begin{document}
 \begin{frontmatter}

 \title
 {Polynomial method for canonical calculations}
 \author[Radium]{N.K. Kuzmenko\corauthref{cor}},
 \corauth[cor]{Corresponding author.} \ead{Kuzmenko@NK9433.spb.edu}
 \author[Univ]{V.M. Mikhajlov}
 \address[Radium]{V.G.Khlopin Radium Institute, 194021
 St.-Petersburg, Russia}
 \address[Univ]{Institute of Physics St.--Petersburg State
 University 198904, Russia}

 \date{\today}
 \begin{abstract}
 A practical version of the polynomial canonical formalism is
 developed for normal mesoscopic systems consisting of $N$ independent
 electrons. Drastic simplification of
 calculations is attained by means of proper ordering excited
 states of the system. In consequence the exact canonical partition
 function can be represented as a series in which the first term
 corresponds to the ground state whereas successive groups of
 terms belong to many particle-hole excitations (~one particle-hole
  two particle-hole and so on). At small
 temperatures ($T<10$ inter-level spacings near the Fermi level)
 the number of terms which should be taken into account is
 weakly dependent on  $N$ and remains $<10$ even if $N\sim 10^5$.
 The elaborated method makes canonical calculations to be not more
complicated than the grand canonical ones and is free from any
limitations on $N$ and $T$.
\end{abstract}

 \end {frontmatter}

 \newpage

 \section{Introduction}
 The characteristic feature of the modern stage in studying
 mesoscopic systems is a possibility to fix individual properties
 of investigated objects such as chemical composition, size geometric
 shape and particle number. The well known example of such
 investigations is the discovery of electronic shells in alkali
 metal clusters~\cite{knight}. The other branch of researches is
  connected with recent advances in the development of new
  techniques for fabricating two- or three-dimensional
 micro- and nano-structures that enables current experiments
 to investigate variations of both
 the exact number of electrons on such mesoscopic structures and
 their geometrical shapes with a precise control in all dimensions.
Frequently experiments are produced so that investigated systems
may exchange energy with surroundings (they are embedded in heat
reservoir), i.e. they are kept at an invariable temperature,
however the exchange by particle is absent. Thus, such systems
constitute a canonical ensemble, and the most appropriate
theoretical method for their statistical description  is the
canonical formalism.

 The papers of
Denton, M\"uhlschlegel and Scalapino~\cite{denton} were the first
where the canonical description was developed in applying to
electrons in small metallic grains. At that time it was impossible
to fix the shape and particle number of each grain. Therefore for
constructing the canonical partition function Denton et al made
the simplest assumption that the energy spectrum of free electrons
in a metallic grain is equidistant ( the equal level spacing
model) and two-fold degenerated if there is no magnetic field.
Using reasonable approximations ($M>N\gg 1$, $M$ is number of
single electron levels) Denton et al established an analytical
expression for the canonical partition function. They showed that
the main differences between the canonical and grand canonical
values of the heat capacity ($C$) and  magnetic susceptibility
($\chi$) lie at small temperatures $T<\delta$ ($\delta$ is the
mean level spacing). Besides they found that at $T\gg\delta$ the
grand canonical heat capacity exceeds the canonical one by $k_B/2$
($k_B$ is the Boltzmann constant).  The partition function of
Denton et al takes into account energy variations in an applied
uniform magnetic field. They manifest themselves in oscillations
of $\chi$ and $C$ with growth of the field. Such oscillations are
typical for charged particle systems ~\cite{kuzmenko3} however the
strong periodicity in the oscillations is a feature of the equal
level spacing.

It is worthwhile emphasizing the $N$-independence of thermodynamic
quantities as functions of $T/\delta$ in the canonical
calculations of Denton et al at $M>N$,  $M,N\gg 1$. It is caused
by a relatively narrow layer of single particle levels near the
Fermi energy, which mainly contributes to the partition function,
and by the identical for any $N$ structure of the single-particle
energy spectrum in this layer.

The canonical formalism was also used by Brack, Genzken and
Hansen~\cite{brack} for calculations of thermal properties of the
valence electrons in alkali metal clusters. In this case free
electron energy spectra contain considerable gaps that accounts
for enhanced stability of clusters with the magic numbers of
atoms. Brack et al showed that the shell structure reveals itself
in the theoretical heat capacity. However, the difference between
the canonical and grand canonical shell effects is quite
noticeable only at temperatures which are less than or of the same
order as the mean level spacing $\delta_F$ near the Fermi. Thus,
both investigations,~\cite{denton,brack}, point out the region of
the temperatures, $T<\delta_F$, where the comparison of the
canonical and grand canonical calculations is of the most
interest. Besides, in the center of attention should be magic
electron numbers (if they appear in the system) since the
existence of such $N$ is a consequence of the appearance of an
energy gap near the Fermi level that enhances the difference
between canonical and grand canonical results. It is well known
that $T<\delta_F$ is the best regime for observing size effects in
mesoscopic systems~\cite{halperin,perenboom}. Therefore the
canonical approach at these temperatures can give more precise
theoretical information than the grand canonical calculations.

In order to calculate the partition function $Z(N,M)$, occupation
numbers, internal energy and other thermodynamic quantities Brack
et al used recurrence relations connecting $Z(N,M)$ with those for
smaller $N$ and $M$. Following this procedure they found that
$N(M-N)$ calculation steps were needed to obtain the final result,
i.e. for $M>N\sim 10^2$ ($N(M-N)\sim 10^4$). Such method is still
manageable for $N$ up to a few hundreds, however it could be
hardly applied to $N$ and $M>10^3$. Thus the constructing of the
effectively working canonical formalism without restrictions on
the particle number and applicable to arbitrary spacing
distributions is a problem that has to be solved.

Parallel with the exact canonical calculations in
Ref.~\cite{denton,brack} approximate methods have been suggested
for constructing the canonical partition function of the normal
(nonsuperconducting) systems. One of them Ref.~\cite{lang}
replaces the projection integral over the gauge variable $\varphi$
($0\leq\varphi\leq 2\pi$) by the discrete sum that involves $L$
terms. These terms depend on $2\pi l/L$ (instead of $\varphi$)
where the integer $l$ varies from $0$ to $L-1$. As shown in Sec.1
this method can be viewed as the partial projection. It is the
more precise the higher is the value of $L$. In
Ref.~\cite{frauendorf} this method was employed by Frauendorf and
Pashkevich to calculate sodium cluster shape and free energies.
Another method suggested by Rossignoli consists in the evaluation
of projection integral in the saddle-point approximation. Such
approach turns out to be quite satisfactory in the case of not
very small temperatures and particle numbers as was demonstrated
by calculations of nuclear level densities in
Ref.~\cite{rossignoli}.

Our effective version of the canonical approach is based on the
polynomial representation of the partition function. Last years
such representation is discussed in the
literature~\cite{schmidt1,philippe}, however our practical version
of polynomial calculations was elaborated many years ago for a
projection method applied to the Bardin-Cooper-Schriffer ($BCS$)
function for describing the particle number conserving pairing
correlations in nuclei~\cite{kuzmenko1}.

There is an obvious analogy between the $BCS$ function and grand
canonical partition function for the independent electron model
($IEM$). Both can be represented as products $\prod (1+f_s)$,
where $f_s$ for the $BCS$ function is $(v_s/u_s)a^{+}_sa^{+}_{\bar
s}$ ($v_s$, $u_s$ are the Bogolubov parameters,
$a^{+}_sa^{+}_{\bar s}$ is the operator of the fermion pair
creation in time conjugated states $s$ and $\bar{s}$) while for
the canonical partition function $f_s$ is
$\exp(\varepsilon_s-\lambda)/kT$, $\varepsilon_s$ and $\lambda$
being the energy of a single -particle state $s$ and the chemical
potential respectively. After projection onto a fixed particle
number $N$ both functions gain the form of a $N$-th order
symmetric polynomial $[N]$. This polynomial $[N]$ consists of
$M!/N!(M-N)!$ terms  (as before $M$ is the total amount of
employed single-particle states) and each term in it is the
production of $N$ different $f_s$, i.e. $[N]$ includes all
distributions of $N$ $BCS$ pairs or $N$ electrons over $M$ states.
For large $M$ and $N$ the total number of terms in $N$ may be
enormous, however a hierarchy of terms can be established
according to their surviving at $G\rightarrow 0$ ($G$ is the
pairing strength) or $T\rightarrow 0$. The first term of this
hierarchy corresponds to the independent fermion ground state in
which all single-particle states are occupied up to
$\varepsilon_F$. The second group of terms corresponds to simplest
excitations: in the $BCS$ theory it is a particle pair
$a^+_sa^+_{\bar s}$ above $\varepsilon_F$ and a hole pair
$a_sa_{\bar s}$ beneath $\varepsilon_F$, for free electrons in a
cluster it is one particle-hole excitations. The third group of
terms corresponds to a shift of two $BCS$ particle pairs or two
cluster electrons above $\varepsilon_F$. This alignment
of many
particle-hole excitations continues till all pairs or cluster
electrons are lifted above $\varepsilon_F$.

The common property of the projected $BCS$ and the canonical
partition functions is mutual independence of probabilities  of
particle and hole excitations. This probability depends only on
the energy of the single-particle level on which a $BCS$ fermion
pair or a cluster electron is located. For the $BCS$ case it is
$(v_s/u_s)$ for levels with $\varepsilon_s >\varepsilon_F$ while
for levels with $\varepsilon_s\leq\varepsilon_F$ it is
$(u_s/v_s)$~, for cluster electrons it is
$\exp(\lambda-\varepsilon_s/kT$ if $\varepsilon_s>\varepsilon_F$,
or $\exp(\varepsilon_s-\lambda)/kT$ if
$\varepsilon_s\leq\varepsilon_F$.
 Thus summing over particle and hole excitations can be performed
 independently. The contribution of each group of $n$-particle
 excitations to $[N]$ includes a sum running distributions of $n$
 electrons (here we mean cluster electrons) over particle levels
 (with energies above $\varepsilon_F$). These sums can be also
 represented as symmetric polynomials, now they are of $n$-th
 order ($n<N$). Analogous polynomials arise for hole excitations,
 but in this case electrons are distributed over hole levels
 (below and including $\varepsilon_F$).
 Though the convergence of appearing series is absolute
 its rate is  determined by $T/\delta_F$.
 Practically for $T/\delta_F<1$ they
 include not more than 5 components (i.e. up to 5 particle-5 hole
  excitations). Our calculations for the pairing
  problem  have indicated high effectiveness of the polynomial
  representation of the  $N$-projected $BCS$-function~\cite{kuzmenko1}
 and this paper is devoted to the adaptation of the polynomial
representation to the canonical partition function of normal
mesoscopic systems.

 The material is arranged as following. In Sec.2 separation of the
full configuration space into two parts, inside and outside the
Fermi sphere, is realized to construct the canonical partition
function $Z_N$ as a symmetrical polynomial of $N$-th order. $Z_N$
is expanded in polynomials of lower orders. These series are
composed of products of ''particle`` and ''hole`` polynomials, the
order of which points out how many particles and holes take part
in excitations allowed for in $Z_N$. In Sec.3 these polynomials
are employed to one- and two-body density matrices and on this
base the canonical expressions for the heat capacity and magnetic
susceptibility are compared to the grand canonical ones. Sec.4
collects recurrent relations for polynomials we apply in
calculations. Sec.5 shows that in a wide range of temperatures and
particle numbers the elaborated method possesses high convergence
and can be used for various types of single-electron energy
spectra. The conclusion is given in Sec.6. Algebraic details for
the equal level spacing model and the exact canonical partition
function of the model are considered in Appendix.

 \section{''Particle`` and ''hole`` symmetric polynomials and
 the canonical partition function}\label{polynom}

 If a system is composed of a fixed number $N$ of constituent
 particles and possesses a discrete excitation spectrum the
 thermodynamic properties of the canonical ensemble ($CE$) of
 such systems can be described by using the partition function
 $Z_N$

 \begin{equation}
 \label{ZN}
 Z_N=\sum_{\alpha} exp \{-\beta E_{\alpha}(N)\},\;\;\;\;
 \beta=T^{-1}
 \end{equation}                                          
 The index $\alpha$ labels energy states of this system
 $E_{\alpha}(N)$, $\alpha=0$ concerns the ground state. The
 calculation of $E_{\alpha}(N)$ for many body systems is a
 complicated problem and in this paper we limit ourselves by the
 $IEM$ which in many cases is a good
 start approximation. Then, assuming that a set of single
 electron energy levels is established for a static mean field
 governing the electron movement, the values of $E_{\alpha}$ are
 readily found in $IEM$.

The ground state energy $E_{\alpha}(N)$ is the sum of $N$ single-
electron energies up to the Fermi energy ($\varepsilon_F$) and the
excited state energies are obtained by addition to $E_{0}$ of
several single-electron particle energies
$\varepsilon_p>\varepsilon_F$) and by subtraction of the same
amount (as $N$ is fixed) of hole energies
($\varepsilon_h\leq\varepsilon_F$)
 \begin{equation}
 \label{E0N}
 E_0(N)=\sum_{s=0}^{F}(\varepsilon_s-\lambda),
 \end{equation}
  \begin{eqnarray}
 \label{Ealfa} E_{\alpha}(N)\equiv E{p_1p_2\ldots p_k,h_1h_2\ldots
 h_k}= E_0(N)+\varepsilon_{p_1}+\varepsilon_{p_2}+\ldots +
 \varepsilon_{p_k}-\nonumber \\
 -\varepsilon_{h_1}-\varepsilon_{h_2}-\ldots - \varepsilon_{h_k}.
 \end{eqnarray}                                           
 We shall treat $\lambda$ in Eq.~(\ref{E0N}) as a chemical
 potential, however, in contrast to the grand canonical ensemble
($GCE$) where $\lambda$ is calculated to allow for the particle
number conservation on the average, the choice of $\lambda$ for
$CE$ as a certain level for counting energies is merely a matter
of convenience (in our calculations $\lambda$ is taken as the
grand canonical chemical potential at given $T$). Substituting
Eqs.~(\ref{E0N},\ref{Ealfa}) into Eq.~(\ref{ZN}) one can see that
the canonical partition function $Z_N$ is a symmetric polynomial
in $q_s$

 \begin{equation}
 \label{Eq4}
 Z_N=\left[ N\right]=\sum_{ \{ distr. \} }\prod_{i=1}^{N} (q_s)_i,
 \;\;\;\; q_s=exp[-\beta (\varepsilon_s -\lambda )].
 \end{equation}                                                 
 The symbol ($\{distr. \}$ in Eq~(\ref{Eq4}) implies that all
 $N$-electron distributions over all single electron levels have to be
 taken into account.

 Another way to introduce $Z_N$ as a symmetrical polynomial consists
 in employing the statistical operator $D_N$ projected onto a fixed
 particle number $N$:

 \begin{eqnarray}
 \label{Eq5}
 \hat{D}_N=\frac{1}{Z_N} P_N\prod_{s} exp \{
 -\beta (\varepsilon_s -\lambda )a^+_s a_s \}, \\
 P_N=\frac{1}{2\pi}\int_{0}^{2\pi} \; d\varphi \;
 exp\{ i\varphi (N-\hat{N} )\}, \; \; \; \;
 \hat{N} =\sum_{s} a^+_s a_s. \nonumber
 \end{eqnarray}                                         
 $a^+_s$, $a_s$ are the creation and annihilation electron operators.
 The normalization of $\hat{D}_N$, $Z_N$, is calculated with unrestricted
 electron basis $(Fock$ $space)\mid\rangle$
 \begin{equation}
 \label{Eq6}
 Z_N=\frac{1}{2\pi}\int_{0}^{2\pi}  \; d\varphi \;
 e^{i\varphi N} \; Z(\varphi ) ;
 \end{equation}                                              

 \begin{eqnarray}\label{Eq7}
 Z(\varphi )=\langle\mid\prod_{s} exp\{ [-\beta (\varepsilon_s
 -\lambda )-i\varphi ]a^+_sa_s \} \mid\rangle = \nonumber \\
= \prod_{s} \{ 1+exp[  -\beta (\varepsilon_s -\lambda )-i\varphi ]
 \} .
 \end{eqnarray}                                      

 The Fourier series for $Z_N$ in powers of $e^{-i\varphi}$ includes  as
 amplitudes symmetrical polynomials $[n]$ of the same type as in
 Eq.~(\ref{Eq4})
 \begin{eqnarray}\label{Eq8}
 Z(\varphi )=1+\sum_{n=1}[n] e^{-i n \varphi }; \\
\label{Eq9} \left[ 1\right] = \sum_{s} q_s; \; \;
 \left[ 2\right] = \sum_{s<t} q_s q_t; \; \;
 \left[ n\right] = \sum_{ \{ distr. \} }\prod_{i=1}^{n} (q_s)_i
 \end{eqnarray}                                               
 The projection in Eq.~(\ref{Eq6}) isolates the polynomial
 with $n=N$ that again leads to Eq.~(\ref{Eq4}).

 If integrating in Eq.~(\ref{Eq6}) ($Z(\varphi)$ from
 Eq.~(\ref{Eq8})) is replaced by summing over discrete variable
 $2\pi l/L$ as was suggested in Ref.~\cite{lang}
\begin{eqnarray}
\frac{1}{2\pi}\int_0^{2\pi} d\varphi e^{i\varphi
(N-n)}\rightarrow\frac{1}{L}\sum_{l=0}^{L-1}e^{i2\pi l(N-n)/L}=
\nonumber\\=\frac{1}{L}\left [1-e^{-L\varepsilon}\right ] \left
\{1- e^{i2\pi l(N-n)/L-\varepsilon)}\right
\}\mid_{\varepsilon\rightarrow 0}=\nonumber\\
 =\left \{  \begin{array}{ll} 1,
&\mbox{$\mid N-n\mid=kL;\;\;\;$ $k=0,1,\ldots$}\\ 0,
 &\mbox{$\mid N-n\mid=kL+\nu;\;\;\;$ $l\leq\nu\leq L-1$},
\end{array}\label{lang}
\right.
\end{eqnarray}
then the number projection does not eliminate completely
components in the grand canonical $Z$ which correspond to $n\neq
N$. However the more is $L$ the larger is the difference between
$N$ and such $n^,$s (not less than $L$).

 The expectation values of the one- or two body density matrices can be
 also expressed by means of polynomials
 \begin{eqnarray}
 \hat{n}_s=a^+_sa_s, \nonumber \\
n_s=\langle \mid\hat{n}_sD_N\mid\rangle =q_s\frac{[N-1]_s}{[N]},
\label{Eqns} \\             
n_{st}=\langle \mid\hat{n}_s\hat{n}_tD_N\mid\rangle =
 q_sq_t\frac{[N-2]_{s,t}}{[N]}, \; \; s\neq t.\label{Eqnst}
 \end{eqnarray}
 If state $s$ is degenerated and can be occupied by several
 electrons, in polynomials there appear powers of the corresponding
 amplitude $q_s$. For such cases we introduce $n_{ss}$:
\begin{equation}
\label{Eqnss}
 n_{ss}=\langle \mid\hat{n}_s\hat{n}_tD_N\mid\rangle
 |_{\varepsilon_t=\varepsilon_s} =
 q_s^2\frac{[N-2]_{s,s}}{[N]}.
 \end{equation}
 Henceforth the notation $[N]_{a,b\ldots }$ implies the exception
 of single electron states $a,b\ldots$ out of the full set of states over
 which $n$ particles are distributed.

 For comparison we write out here the $GCE$ results for the same
 quantities:
 \begin{equation}
 \label{Eq12}
 Z(GCE)=\prod_s (1+q_s)=1+\sum_{n=1}[n].
 \end{equation}
 The maximum $n$ in Eq.~(\ref{Eq12}) is determined by the amount of
 the single electron states under consideration.
 \begin{eqnarray}
 \label{Eqfs}
 \langle \mid\hat{n}_sD(GCE)\mid\rangle =f_s=1/(1+q_s^{-1}),\\
 \langle \mid\hat{n}_s\hat{n}_td(GCE)\mid\rangle =
 f_sf_t, \label{Eqfsft} \; \; s\neq t,
 \end{eqnarray}
 $f_s$ is the Fermi occupation numbers.

Eq.~(\ref{Eqnst}) shows that for the canonical ensemble the
two-body density matrix $n_{st}$ is not reduced to a product of
two one-body ones $n_sn_t$ as it takes place in $GCE$,
Eq.~(\ref{Eqfsft}). Thereby the canonical description allows for
correlations in electron movement in spite of the starting
assumption of the $IEM$.

 The method we are about to develop here for computations of
 polynomials entering into
 Eqs.~(\ref{Eq4}),~(\ref{Eqns})-(\ref{Eqnss}) takes advantage of the
 general property of symmetrical polynomials: each polynomial
 given in a space of variables $q_s$ can be represented as a
 decomposition of polynomials of the same and lower orders defined
 in subspaces of smaller dimensions. That can be viewed as a
 consequence of the Clebsch-Gordan decomposition applied to the
 basic vectors of the symmetric group irreducible representations.
 If the original space $C$ spans a set of single electron states
 which can be divided into two subsets ($C=A\oplus B$) in
 accordance with energy or any other quantum numbers then an
 initial polynomial $\left [n\right ]^{(C)}$ can be represented as
 following
 \begin{equation}
 \label{Eq15}
 \left [n\right ]^{(C)}=\sum_{\nu=0}^{n}\left [n-\nu\right ]^{(A)}
 \left [\nu\right ]^{(B)};\;\;\;\;
 \left [0\right ]^{(A)}=\left [0\right ]^{(B)}=1.
 \end{equation}
 Superscripts in this equation denote the spaces in which polynomials are
 defined. The order of each polynomial $\left [n-\nu\right ]^{(A)}$ or
 $\left [\nu\right ]^{(B)}$ in Eq.~(\ref{Eq15}) cannot exceed the
 dimension of space $A$ or $B$ respectively.

 The first realization of this decomposition is dividing the whole
 single electron energy space into two parts: below and above the Fermi
 level ($F$) that drastically simplifies the calculation of the
 polynomials and hence, as it will be shown in Sec.5, enables
 the canonical approach to be employed in a wide range of temperatures
 and particle numbers.

 To implement such dividing we have to explicitly allow for possible
 degeneracy of electron states and in particular the degeneracy of
 the Fermi level,
 $d_F$. If $d_F>1$ the occupation rate of $F$ dictates three ways
 to factorize of $\left [N\right ]$
\begin{equation}\label{EqZ0ZT}
Z_N= \left [N\right ]=Z^{(0)}\widetilde{Z}_N
\end{equation}

The first case corresponds to the completely filled Fermi level at
$T=0$ ($n_F=d_F$, $n_F$ is the particle number on F-level). Then
the first factor on the right side of Eq.~(\ref{EqZ0ZT}) is the
polynomial $\left [N\right ]^{(A)}$ where space $A$ of the
dimension $N$ involves all states below and including $F$.
Therefore this polynomial consists of only one term and is the
product of all $q_s$ with $s$ running states $\leq F$. Factoring
out this term in Eq.~(\ref{EqZ0ZT}) leads to a series in
symmetrical polynomials:
 \begin{eqnarray}\label{Eq17}
 Z^{(0)}=\prod_{s\leq F}(q_s)^{d_s}=exp\left[-\beta
 E_0(N)\right],\;\;\;\;
 \widetilde{Z}_N=\sum_{n=0}^{N}\overline{[n]}[n],\\
 \overline{[0]}=[0]=1, \nonumber
 \end{eqnarray}                                        
 $d_s$ is the degeneracy of state $s$ ($d_s$ is equal to the maximum
 electron number on this level);
 $[n]$ and $\overline{[n]}$ are $n$-th order polynomials, the former
 is a ``particle'' polynomial defined in variables $q_s=exp\left
 [\beta(\lambda-\varepsilon_s)\right ]=f_s/(1-f_s)$,
 $\varepsilon_s>\varepsilon_F$ while the latter is a ``hole''
 polynomial in $q^{-1}_t=exp\left
 [\beta(\varepsilon_t-\lambda)\right ]=(1-f_t)/f_t$,
 $\varepsilon_t\leq\varepsilon_F$, $f_s$ is determined in
 Eq.~(\ref{Eqfs}).
Thus, $\lambda$ being chosen between $F$ and $F+1$ (the level just
above $F$), both $[n]$ and $\overline{[n]}$ are functions of
quantities which are $<1$. This choice of $\lambda$ and
exponential  energy  damping of variables $q_s$ and $q_t^{-1}$
provide the prompt convergence of the series in Eq.~(\ref{Eq17})
especially at small temperatures.

The second case is $n_F\leq d_f/2$ at $T=0$. Now the factors
$Z^{(0)}$ and $\widetilde{Z_N}$  are
 \begin{equation}
 \label{Eq18}
 Z^{(0)}=\prod_{s}^{F-1}(q_s)^{d_s}; \;\;\;
 \widetilde{Z}_N=\sum_{n=0}^{N-n_F}\overline{[n]}[n_F+n],
 \end{equation}                                               
 $F-1$ is the topmost state just below $F$.

 The third case is $d_F/2<n_F<d_F$ at $T=0$, then
 \begin{equation}
 \label{Eq19}
 Z^{(0)}=\prod_{t}^{F}(q_t)^{d_t}; \;\;\;
 \widetilde{Z}_N=\sum_{n=0}^{N}\overline{[d_F-n_F+n]}[n].
 \end{equation}                                              

\section{Canonical magnetic susceptibility and heat capacity}

The polynomials $[n]$, $[{\bar n}]$ and those with omitted states
determine one- and two-body density matrices
(Eqs.~(\ref{Eqns}-\ref{Eqnss})). Their explicit form for the
$n_F=d_F$ case is following ( $\widetilde{Z}_N$ is defined by
Eq.~(\ref{Eq17})):
 \begin{eqnarray} \label{ns}
 n_s=\left \{ \begin{array}{ll}
1-(q_s^{-1}/\widetilde{Z}_N) \sum_{n=1}^{N}[n]\overline{[n-1]}_s,
 & \mbox{  if  $s \leq F$} \\
(q_s/\widetilde{Z}_N)\sum_{n=1}^{N}[n-1]_s\overline{[n]},
 & \mbox{  if $s > F$};
\end{array}
\right.
\end{eqnarray}
\begin{equation}
\label{nst}
 n_{st}=\frac{q_tn_s-q_sn_t}{q_t-q_s}\;\;\;\;\; s\neq t;
\end{equation}
 \begin{eqnarray}
  \label{nss}
 n_{ss}=\left \{ \begin{array}{rr}
1-(2q_s^{-1}/\widetilde{Z}_N)\sum_{n=1}^{N}[n]\overline{[n-1]}_s+
\\
+(q_s^{-2}/\widetilde{Z}_N)\sum_{n=2}^{N}[n]\overline{[n-2]}_{ss},
 & \mbox{ if $s \leq F$} \\
(q_s^2/\widetilde{Z}_N)\sum_{n=2}^{N}[n-2]_{ss}\overline{[n]},
 & \mbox{  if $s > F$}.
\end{array}
\right.
\end{eqnarray}

These density matrices and analogous ones that correspond to
Eqs.~(\ref{Eq18}),(\ref{Eq19}) determine $\chi_{\mathrm{can}}$ and
$C_{\mathrm{can}}$:
\begin{equation}\label{Scan}
\chi_{\mathrm{can}}=\chi^{(P)}_{\mathrm{can}}+\chi^{(D)}_{\mathrm{can}},
\end{equation}
Superscripts $P$ and $D$ mean the paramagnetic and diamagnetic
parts of $\chi$.
\begin{eqnarray}
\chi^{(P)}_{\mathrm{can}}=\frac{\mu_B^2\beta}{V}\left[\sum_s d_s
(\frac{\partial\varepsilon_s}{\partial\omega})^2 n_s(1-n_s)
\nonumber\right. \\ +\sum_s
d_s(d_s-1)(\frac{\partial\varepsilon_s} {\partial\omega})^2
(n_{ss}-n_s^2)\nonumber \\ + 2\sum_{s>t} d_sd_t(
\frac{\partial\varepsilon_s}{\partial\omega})\left.
(\frac{\partial\varepsilon_t}{\partial\omega}) (n_{st}-n_sn_t)
\right],  \label{SPcan}
\\
\chi^{(D)}_{\mathrm{can}}=-\frac{\mu_B^2}{V}\sum_sd_s
\frac{\partial^2\varepsilon_s}{\partial\omega^2}n_s,\label{SDcan}\\
C_{\mathrm{can}}/k_B=\beta^2\left[\sum_s d_s
(\varepsilon_s-\lambda )^2n_s(1-n_s)\right. \nonumber \\ +\sum_s
d_s(d_s-1)(\varepsilon_s -\lambda)^2(n_{ss}-n_s^2)\nonumber
\\ + \left. 2\sum_{s>t}
d_sd_t(\varepsilon_s-\lambda)(\varepsilon_t -\lambda)
(n_{st}-n_sn_t)\right] \label{Ccan2}
\end{eqnarray}
The correlations caused by the particle number conservation
formally reveal themselves in the appearance two in
Eqs.~(\ref{SPcan},\ref{Ccan2}) terms containing two-body density
matrices ($n_{st}$ and $n_{ss}$). Besides, due to the
correlations, i.e. due to the dependence of the occupation
probability of a level on the occupation of the other levels, the
Fermi occupation numbers $f_s$ differ from the canonical ones
$n_s$. The grand canonical  $\chi_{\mathrm{grand}}$ and
$C_{\mathrm{grand}}$ can be obtained from
Eqs.~(\ref{SPcan},\ref{Ccan2}) by replacing $n_s$ by $f_s$ and
removing terms proportional to $(n_{st}-n_sn_t)$ and
$(n_{ss}-n_s^2)$. Both these factors now vanish since in the
absence of the correlations in the grand canonical ensemble
two-body matrices are factorized: $(n_{st}=n_sn_t)$ and
$(n_{ss}=n_s^2)$.
\begin{eqnarray}
\chi_{\mathrm{grand}}=\chi^{(P)}_{\mathrm{grand}}+\chi^{(D)}_{\mathrm{grand}}
\label{Sgrand},\\
\chi^{(P)}_{\mathrm{grand}}=\frac{\mu_B}{V}\beta\sum_s d_s\left
(\frac{\partial\varepsilon_s}{\partial\omega}\right)^2f_s(1-f_s),
\label{SPgrand} \\
\chi^{(D)}_{\mathrm{grand}}=-\frac{\mu_B}{V}\sum_sd_s
\frac{\partial^2\varepsilon_s}{\partial\omega^2}f_s,\label{SDgrand},\\
 C_{\mathrm{grand}}/k_B=\beta^2\sum_s d_s
(\varepsilon_s-\lambda )f_s(1-f_s)(\varepsilon_s-\lambda
-\beta\frac{\partial\lambda}{\partial\beta}), \label{Cgrand1}\\
\frac{\partial\lambda}{\partial\beta}=\frac{\sum d_s
(\varepsilon_s-\lambda )f_s(1-f_s)}{\beta\sum d_s f_s(1-f_s)}
\end{eqnarray}
Eq.~(\ref{Cgrand1}) for $C_{\mathrm{grand}}$ is written at the
condition that the average value of $N$ is conserved. In this case
the chemical potential is subject to temperature variations, term
in Eq.~(\ref{Cgrand1}) $\beta\partial\lambda/\partial\beta$. This
term is absent if measurements are performed at a constant
$\lambda$. An analogous term in the susceptibility is dropped,
since it is vanishingly small.

\section{Calculations of the polynomials}

Polynomials entering into $\widetilde{Z_N}$, $\chi_{\mathrm{can}}$
and $C_{\mathrm{can}}$ can be expressed
 through $q_s$ and $q_t^{-1}$ with help of recurrent procedure
 based on Eq.~(\ref{Eq15}). If in this equation the partition
 $C=A\oplus B$ is such that space $B$ consists of only one state
 $s$ then $[\nu]^{B}=\delta_{\nu.1}q_s+\delta_{\nu.0}$ and
 polynomial $[n-\nu]^{(A)}$ in space $A$  is in fact $[n-\nu]_s$
 since it is defined in the whole space $C$  with state $s$ omitted.
 Eq.~(\ref{Eq15}) can be therefore rewritten as
 \begin{equation}
 \label{Eq20}
 [n]=[n-1]_sq_s+[n]_s.
 \end{equation}                                            
 The sum rule gives the second equation for successive
 calculations
 of polynomials:
 \begin{equation}
 \label{Eq21}
 \frac{1}{n}\sum_{s}[n-1]_sd_sq_s=[n].
 \end{equation}                                              
 The meaning of Eq.~(\ref{Eq21}) is illustrated by its application to
 Eq.(10)
 \begin{equation}
 \label{Eq22}
 \sum_{s}[N-1]_sd_sq_s=[N]\langle\mid\hat{N}D_N\mid\rangle
 =[N]\cdot N.
 \end{equation}                      
 Here $s$ enumerate all single electron states.

 The lowest order polynomials being found straightforwardly
 \begin{equation}
 \label{Eq23}
 [0]=[0]_t=1; \;\;\;
 [1]=\sum_{s}d_sq_s,
 \end{equation}                      
 higher order ones can be determined by using Eq.~(\ref{Eq20}),
 that gives the well-known Newton identity
 (\ref{Eq21})
 \begin{equation}
 \label{Eq24}
 [n]=\frac{1}{n}\sum_{\nu=1}^{n}[n-\nu]a_{\nu}(-1)^{\nu-1},
 \end{equation}                      
 \begin{equation}
 \label{Eq25}
 a_{\nu}=\sum_{s} d_s{q_s}^{\nu}.
 \end{equation}                      
 Eq.~(\ref{Eq24}) can be used to express $[n]$ only through $a_{\nu}$,
 Eq.~(\ref{Eq25}), such formula is given in textbooks on algebra, but
 it is not so convenient for calculations as Eq.~(\ref{Eq24}).

 \begin{equation}
 \label{Eq26}
 [n]_t=\sum_{\nu=0}^{n}[n-\nu]q_t^{\nu}(-1)^{\nu}.
 \end{equation}                        
 If $q_s\ne q_t$ for states $s$ and $t$ then

 \begin{eqnarray}\label{Eq27}
 [n]_{st}=\sum_{\nu=0}^{n}[n-\nu](-1)^{\nu}\sum_{\mu=0}^{\nu}
 q_s^{\nu-\mu}q_t^{\mu}\nonumber \\
 =(q_t[n]_t-q_s[n]_s)/(q_t-q_s)\\
 =([n+1]_t-[n+1]_s)/(q_s-q_t).
 \end{eqnarray}                        
 If state $s$ is degenerated ($d_s>1$) then it is possible
 to define $[n]_{ss}$,  Eq.~(\ref{Eqnss}), as following
 \begin{equation}
 \label{Eq28}
 [n]_{ss}=\sum_{\nu=0}^{n}[n-\nu](-1)^{\nu}q_s^{\nu}(\nu +1)
 \end{equation}                         
 or by using recurrent relations
 \begin{equation}
 \label{Eq29}
 [0]_{ss}=1, \;\;\;\;
 [n]_{ss}=[n]_{s}-q_s[n-1]_{ss}
 \end{equation}                         

 Eqs.~(\ref{Eq24}), (\ref{Eq26}-\ref{Eq28})
 are sums of sign-changing terms. Instead of them one can use sums
 with positive terms
 \begin{equation}
 \label{Eq30}
 [n]^{(C)}=\sum_{\mu_1+\mu_1+\ldots+\mu_n=n}
 [\mu_1]^{(S_1)}\cdot[\mu_2]^{(S_2)}\cdot\ldots\cdot[\mu_n]^{(S_n)}.
 \end{equation}                         
 that can be obtained from Eq.~(\ref{Eq15}). Superscripts in
 Eq.~(\ref{Eq30}) assume that $C=S_1\oplus S_2\oplus\ldots\oplus
 S_n$ where each subspace $S_i$ embraces $d(S_i)$ single electron
 states, therefore $\mu_i\leq d(S_i)$. If these states are
 degenerated, any polynomial in Eq.~(\ref{Eq30}) is calculated
 analytically
 \begin{equation}\label{Eq31}
 [\mu]^{(S)}=C^{d(S)}_{\mu}
  exp \{-\beta\mu (\varepsilon_s -\lambda)\},\;\;\;
 C_{\mu}^{d(S)}=\left(
 \begin{array}{c}
 d(S)  \\
 \mu
 \end{array} \right).
 \end{equation}
 $C^{A}_{B}$ is a binomial coefficient.
 Eqs.~(\ref{Eq30}), (\ref{Eq31}) are appropriate to calculate
 the derivatives of $[n]$ with respect to $q_s$
 \begin{equation}\label{Eq32}
 \frac{\partial [n]}{\partial q_s}=d_s[n-1]_s; \;\;\;
 \frac{\partial^2 [n]}{\partial q_s^2}=d_s(d_s-1)[n-2]_{ss}.
 \end{equation}                                  
 If each subspace $S_i$ is nondegenerate, i.e. each $\mu_i=0$ or $1$,
 Eq.~(\ref{Eq30}) is transformed into the sum over distributions
 similar to Eq.~(\ref{Eq4}).

 When a magnetic field is applied to a system in which spin and
 orbital momenta are decoupled each electron spin is aligned along
 or against the field $B$. That gives rise to the Zeeman splitting
 of the single electron levels
 \begin{equation}\label{Eq33}
 \varepsilon_s=\varepsilon_s(B)\pm\omega, \;\;\;\;
 \omega=\mu^*_B\frac{g}{2}B,
 \end{equation}                                  
 $\mu^*_B$ is the effective Bohr magneton and $g$ is the effective
 gyromagnetic ratio. $\varepsilon_s(B)$ incorporates magnetic
 effects on the spatial electron density. The spin splitting can be
 explicitly separated in $Z_N$ by means of dividing the whole space
 $C$ of single electron states into two subspaces: $C=C^{(+)}\oplus
 C^{(-)}$, $C^{(+)}$ spans states with positive magnetic spin
 contributions $+\omega$ to the energy, Eq.~(\ref{Eq30}),
 $C^{(-)}$ corresponds to $-\omega$. Such partition causes the
 decomposition of each polynomial:
 \begin{equation}
 \label{Eq34}
 [n]=\sum_{\nu=0}^{\nu_m}
 [n-\nu]^{(+)}[\nu]^{(-)}; \;\;\;
 \nu_m=min(n,G).
 \end{equation}                         
 $G$ is the whole amount of single particle levels which were twice
 degenerate in the absence of the field. Polynomials $[\nu]^{(\pm)}$
 include factors $exp[\mp\beta\omega\nu]$ that allows one to reduce
 Eq.~(\ref{Eq34}) to the form
 \begin{eqnarray}\label{Eq35}
 [n]=\exp\{-\beta\omega n\}\sum_{\nu=0}^{n}
 \exp\{2\beta\omega\nu \}[[n-\nu]][[\nu]]=\nonumber \\
 \sum_{\nu=0}^{(n-\xi_n)/2}[2-(1-\xi_n)\delta_{\nu.0}]\left[
 \left[
 \frac{n+\xi_n}{2}+\nu\right]\right]
 \left[\left[ \frac{n-\xi_n}{2}-\nu\right]\right]\times
 \nonumber \\
 \cosh\left[ \beta\omega(2\nu+\xi_n\right]\, , \\
 \xi_n=[1-(-)^n]/2\, . \nonumber
 \end{eqnarray}                        
The polynomials $[[(n\pm \xi_n)/2\pm\nu]]$ in Eq.~(\ref{Eq35}) are
composed of $q_s$ independent of the spin magnetism: $q_s=\exp
\{-\beta[\varepsilon_s(B)-\lambda]\}$, $\varepsilon_s(B)$ is given
by Eq.~(\ref{Eq33}). Eq.~(\ref{Eq35}) is employed in Appendix to
find $Z_N$ for the equal level spacing model.
Eqs.~(\ref{Eq21}~-~\ref{Eq35}) are written out for ``particle''
polynomials $[n]$ for which space $G$ stretches from level $(F+1)$
up to the topmost level $M$. The same equations are applied to
``hole'' polynomials $\overline{[n]}$ provided variables $q_s$
with $(\varepsilon_s(B)-\lambda)>0$ are replaced by $q_t^{-1}$
with $(\varepsilon_t(B)-\lambda)<0$. The level space for $[{\bar
n}]$ ranges along all single-particle space beneath and including
level $F$.

\section{The potentialities of the polynomial method}

As mentioned in Introduction the best temperature regime to reveal
the difference between canonical and grand canonical description
is $T<\delta_F$. At this condition the convergence of series over
$n$ for $\widetilde{Z}_{N}$,  Eqs.~(\ref{Eq17})-(\ref{Eq19}), is
attained by taking into account as low as a few terms ($n\ll N$).
However, in this section we will show that the polynomial method
can be applied in a wide temperature range since these series
possess the absolute convergence. Moreover, even for such high
temperature as $T\sim 50\delta_F$ real values of $n$ do not
surpass $n_{\mathrm{max}}\sim 100$ for $N\sim 10^5$, i.e.
$n_{\mathrm{max}}$ is far smaller than $N$, and the most part of
series for $\widetilde{Z}_{N}$, which in the general case stretch
to $N$, does not contribute to $\widetilde{Z}_{N}$.

Firstly we shall accomplish estimates of the convergence
analytically and exploit for this purpose the equal level spacing
model that is reasonable approximation to a real system at high
temperatures $T>\delta_F$ when details of possible shell structure
are averaged out. To avoid the double degeneration of levels we
dispose the system in a weak magnetic field $B$ to actualize the
Zeeman splitting, Eq.~(\ref{Eq33}). If $\omega=\Delta/4$ ($g_s=2)$
the levels are equidistantly placed as before with the level
spacing $\Delta/2$. Here and in Appendix $\Delta$ is used for the
level spacing at $B=0$. In this case the polynomials entering into
$\widetilde{Z}_{N}$
\begin{equation}\label{zn}
\widetilde{Z}_{N}=\sum_{n=0}^Nz(n),
 \;\;\;\; z(n)=[n]\overline{[n]},
 \end{equation}
 are defined in a space of ''particle``
($q_s$) and ''hole`` ($q_t^{-1}$) variables, Eq.~(\ref{Eq4}),
reduced to powers of $q$ (here we put $\lambda=\varepsilon_F$):
\begin{eqnarray}
\label{q} q=\exp(-1/t),\;\;\;\; t=2T/\Delta \\
q_s=\exp\left[-\beta(\varepsilon_{F+s}-\varepsilon_F\right]=q^s\\
q_t^{-1}=\exp\left[-\beta(\varepsilon_F-\varepsilon_{F-t})\right]=q^t
\end{eqnarray}                   
As shown in Appendix polynomials in such variables can be found
analytically, Eq.(A3). If $M>N\gg 1$ these expressions are
simplified
\begin{equation}
\label{eqnn} \overline{[n]}=
q^{n(n-1)/2}\prod_{\nu=1}^{n}(1-q^{\nu})^{-1},\;\;\;\;
[n]=q^n\overline{[n]}.
\end{equation}

\begin{figure}[p]
\scalebox{0.5}{\includegraphics{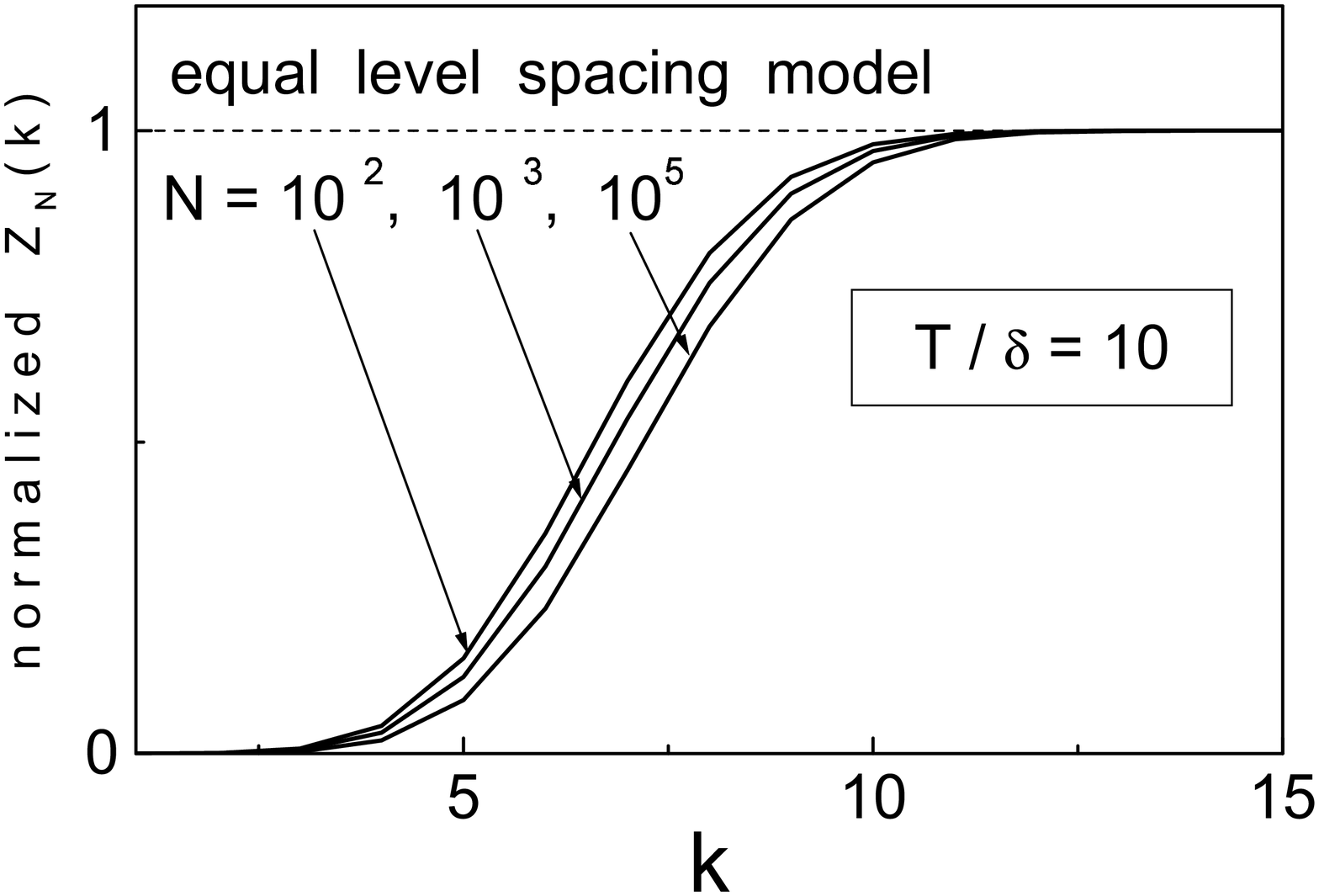}}
\caption{{\label{NormN}} The rate of the convergence of series in
symmetrical polynomials. Normalized $Z_N(k)$ is
$\sum_{n=0}^k[n][\bar n]/\sum_{n=0}^{N}[n][\bar n]$.}
\end{figure}

At small $T$ when $t\ll 1$ the series for $\widetilde{Z}_{N}$ is
exhausted by few first terms since $q\ll 1$. However the
convergence falls if $T$ increases giving rise to $q\sim 1$. At
such temperatures the polynomials $[n]$ are increased with $n$ up
to some $n_0$ after which they begin decreasing. An estimation of
$n_0$ is obtained from the condition:
\begin{equation}
\label{condn0} \frac{z(n_0-1)}{z(n_0)}<1,
 \;\;\;\frac{z(n_0+1)}{z(n_0)}<1,
\end{equation}                         
Substituting  Eq.~(\ref{eqnn}) into Eq.~(\ref{condn0}) gives
\begin{equation}
\label{n0} n_0\simeq t\ln(1+\exp\{1/2t\})\simeq 0.7t.
\end{equation}                   
Thus $n_0$ increases with $t$ and practically does not depend on
$N$ if $n_0\ll N$. The upper limit of summation,
$n_{\mathrm{max}}$, can be found from the condition
\begin{displaymath}
\frac{z(n_{\mathrm{max}}+1)}{z(n_{\mathrm{max}})}< 0.1
\end{displaymath}
This equation yields $n_{\mathrm{max}}\sim 2n_0$ that is
demonstrated by Fig.~\ref{NormN} where the results of the exact
calculations of $\widetilde{Z}_{N}$ for the equal level spacing
model are given. It displays the ratios of $\widetilde{Z}_{N}$
calculated with few terms of the full sum in  Eq.~(\ref{zn}) to
the exact value of $\widetilde{Z}_{N}$. At a fixed temperature the
bend points for $N=10^2\div 10^5$ practically coincide. These
points correspond to $n_0$ and the saturation sets in even earlier
than at $2n_0$.

\begin{figure}[p]
\scalebox{0.65}{\includegraphics{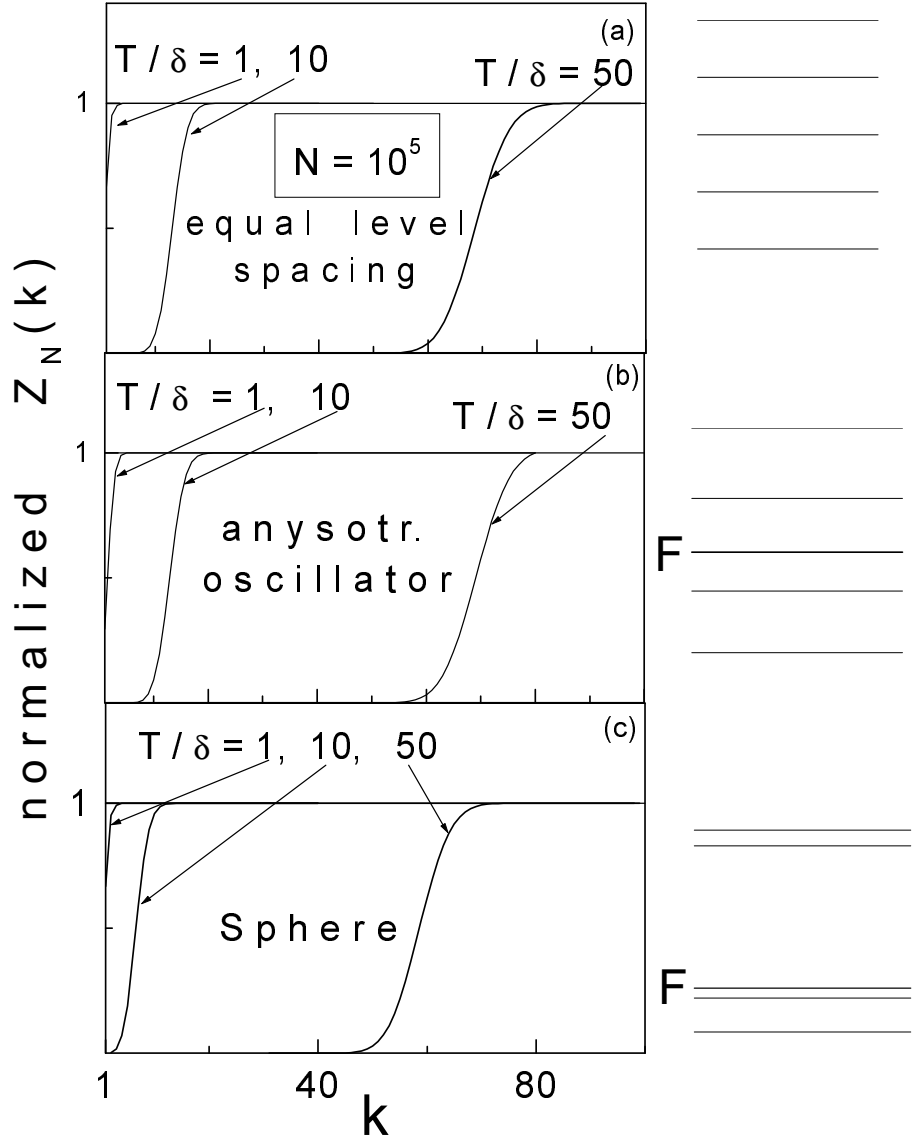}}
\caption{{\label{NormT}}Influence of the temperature on the
convergence of series $Z_N$ for different model spaces. $Z_N(k)$
is defined in the capture to Fig.1. $\delta$ is the mean level
spacing near the Fermi level. The right panel shows fragments the
single-electron spectra near the Fermi level for these model
spaces:(a) the equal level spacing; (b) an anysotropic oscillator
potential with frequencies $\omega_x/\omega_y=1.33$,
$\omega_x/\omega_z=1.55$; (c) a spherical cavity with
$R=r_0N^{1/3}$.}
\end{figure}

As shown in  Fig.~\ref{NormT} the values of $n_0$ at a fixed
reduced temperature ( $T/\delta_F$) depends on the type of the
electron confinement. The energy single-electron spectra of the
equal level spacing model and that of the anysotropic oscillator
are similar. Therefore similar is the convergence of the
corresponding $\widetilde{Z}_{N}$-series. For a spherical cavity
the spectrum reveals shell gaps. In consequence the convergence
turns out to be somewhat higher.

At high temperatures $q\simeq 1-1/t$ and $1-q^n\simeq n/t$,
therefore the $n$-dependence of any term $z(n)$ in  Eq.~(\ref{zn})
at $n\gg 1$ can be obtained from  Eq.~(\ref{eqnn})
\begin{equation}\label{EqZn}
z(n)\mid_{n\gg 1} \sim\frac{(t)^{2n}}{(n!)^2},
\end{equation}                             
that proves the absolute convergence of
$\widetilde{Z}_{N}$-series.

Asymptotic values of $\widetilde{Z}_{N}$ at high temperatures can
be found if summing over single-particle states is replaced by
integrating with that part of the level density
$\rho(\varepsilon)$ which is monotonously varying with energy
$\varepsilon$. At such $T$ the fluctuation part of
$\rho(\varepsilon)$ (shell correction) practically brings no
contribution. Along with this approximation two other conditions
are employed: $T\ll \varepsilon_F$ and $N\gg 1$.

$Z_{N}$ in $IEM$ defined as an integral, Eqs.(6),(7), can be
represented as following
\begin{eqnarray}\label{intZn}
Z_{N}=\exp\left\{-\beta(E_0-\lambda N)\right\}
\frac{1}{2\pi}\int_0^{2\pi}d\varphi\exp(A+B)\\ \label{A}
A=\sum_{s\leq F}\ln\left\{
1+\exp\right[\beta(\varepsilon_s-\lambda)
 +i\varphi\left]\right\}\rightarrow \nonumber\\
\rightarrow \beta^{-1} \int_0^{\lambda}\rho(\lambda
 -y\beta^{-1})\ln\left[1+\exp(-y-i\varphi)\right] dy\\
 \label{B} B=\sum_{t>F}
 \ln\left\{ 1+\exp\right[-\beta(\varepsilon_t-\lambda)
 +i\varphi\left]\right\}\rightarrow \nonumber \\
\rightarrow\beta^{-1} \int_0^{\infty}\rho(\lambda
 -y\beta^{-1})\ln\left[1+\exp(-y-i\varphi)\right] dy
\end{eqnarray}                             
For $N\gg 1$ the upper limit $\lambda$ in  Eq.~(\ref{A}) is
extended to $\infty$. Then taking into account only first term
(independent of $\beta$) in the expansion of functions
$\rho(\lambda\pm y\beta^{-1})$ one arrives at
\begin{eqnarray}\label{A+B}
A+B=\rho(\lambda)\beta^{-1}\int_{0}^{\infty}\ln\left(
1+2\cos\varphi e^{-y}+e^{-2y}\right)dy=\nonumber \\
  =\frac{\rho(\lambda)\beta^{-1}}{2}(\pi^2-\varphi^2),\;\;\;\;
\varphi\leq\pi,
\end{eqnarray}                       
Thus $Z_{N}$ gains the form
\begin{eqnarray}\label{intZn1}
Z_{N}=\exp\left\{-\beta(E_0-\lambda
N)+\pi^2\rho(\lambda)/6\beta\right\}\xi,\nonumber
\\
\xi=\frac{1}{\pi}\int_0^{\pi}d\varphi\exp\left[-\rho(\lambda)\beta^{-1}
\varphi^2/2\right]
\end{eqnarray}

As $\rho(\lambda)\beta^{-1}$ is supposed to be $\gg 1$ the upper
limit in the integral in  Eq.~(\ref{intZn1}) can be taken as
$\infty$:
\begin{equation}\label{intZn2}
Z_{N}=\left(\frac{\beta}{2\pi\rho(\lambda)}\right)^{1/2}
\exp\left[-\beta(E_0-\lambda N)+\pi^2\rho(\lambda)/6\beta\right]
\end{equation}
The value of $Z_{\mathrm{grand}}$ is found from Eq.~(\ref{intZn1})
if $\xi=1$
\begin{equation}\label{intZgrand}
Z_{\mathrm{grand}}=\exp\left[-\beta(E_0-\lambda
N)+\pi^2\rho(\lambda)/6\beta\right].
\end{equation}
In  Eq.~(\ref{intZgrand}) $\lambda$ is regarded as a function of
$T$ defined by the particle number conservation.

Now one can see the difference between $C_{\mathrm{can}}$ and
$C_{\mathrm{grand}}$
\begin{equation}\label{Ccan1}
C_{\mathrm{can}}=k_B\beta^2\frac{\partial^2\ln Z_{N}}
{\partial\beta^2}=
 k_B^2T\rho(\lambda)\pi^2/3-k_B/2;
\end{equation}
\begin{eqnarray}
C_{\mathrm{grand}}=k_B\beta^2\frac{\partial U}{\partial\beta};
\nonumber\\ U=-\frac{\partial\ln
Z_{\mathrm{grand}}}{\partial\beta}
\mid_{\lambda=const}+\lambda\beta^{-1}\frac{\partial\ln
Z_{\mathrm{grand}}}{\partial\lambda},\nonumber \\
 \label{Cgrand}C_{\mathrm{grand}}=k_B^2T\rho(\lambda)\pi^2/3.
\end{eqnarray}                     
In Eq.~(\ref{Cgrand}) the condition $T/\varepsilon_F\ll 1$ and the
relation
\begin{displaymath}
\rho(\lambda)\partial\lambda/\partial\beta\simeq\beta^{-3}
(\partial\rho/\partial\lambda)\pi^2/3
\end{displaymath}
 are employed. Thus the
difference between $C_{\mathrm{can}}$ and $C_{\mathrm{grand}}$ is
the same as it was found in Ref.~\cite{denton} for the equal level
spacing model.
 Eqs.~(\ref{Ccan1},\ref{Cgrand}) show that the same
 difference exists for any level density.

\begin{figure}[p]
\scalebox{0.5}{\includegraphics{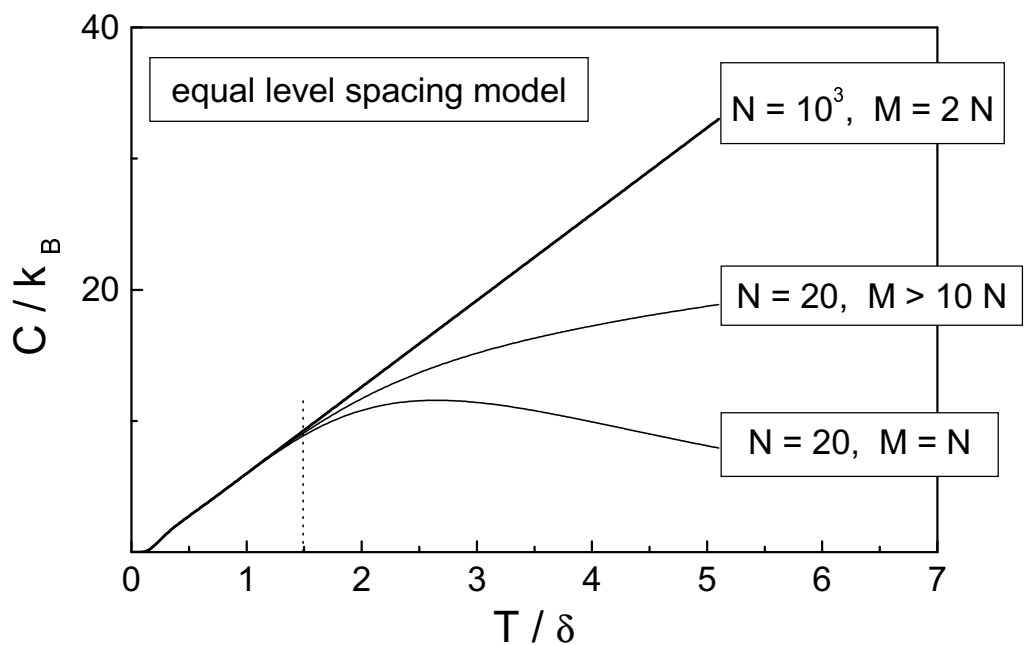}}
\caption{{\label{mdepend}} The canonical heat capacity vs
temperature. The bold curve represents the results of
Ref.~\cite{denton} and coincide with the exact canonical results
if  $N\geq 10^3$. For smaller $N$ the high temperature
($T>1.5\delta$) behavior of $C_{\mathrm{can}}$ (the curve for
$N=20$, $M\geq 10N$, $M$ is the number of spin degenerate
single-electron levels) differs from predictions of
Ref.~\cite{denton}. The curve for $N=20$, $N=M$ shows that high
temperature behavior of $C$ in small electron ensembles depends on
$M$ if $M<10N$. }
\end{figure}

\begin{figure}[p]
\scalebox{0.5}{\includegraphics{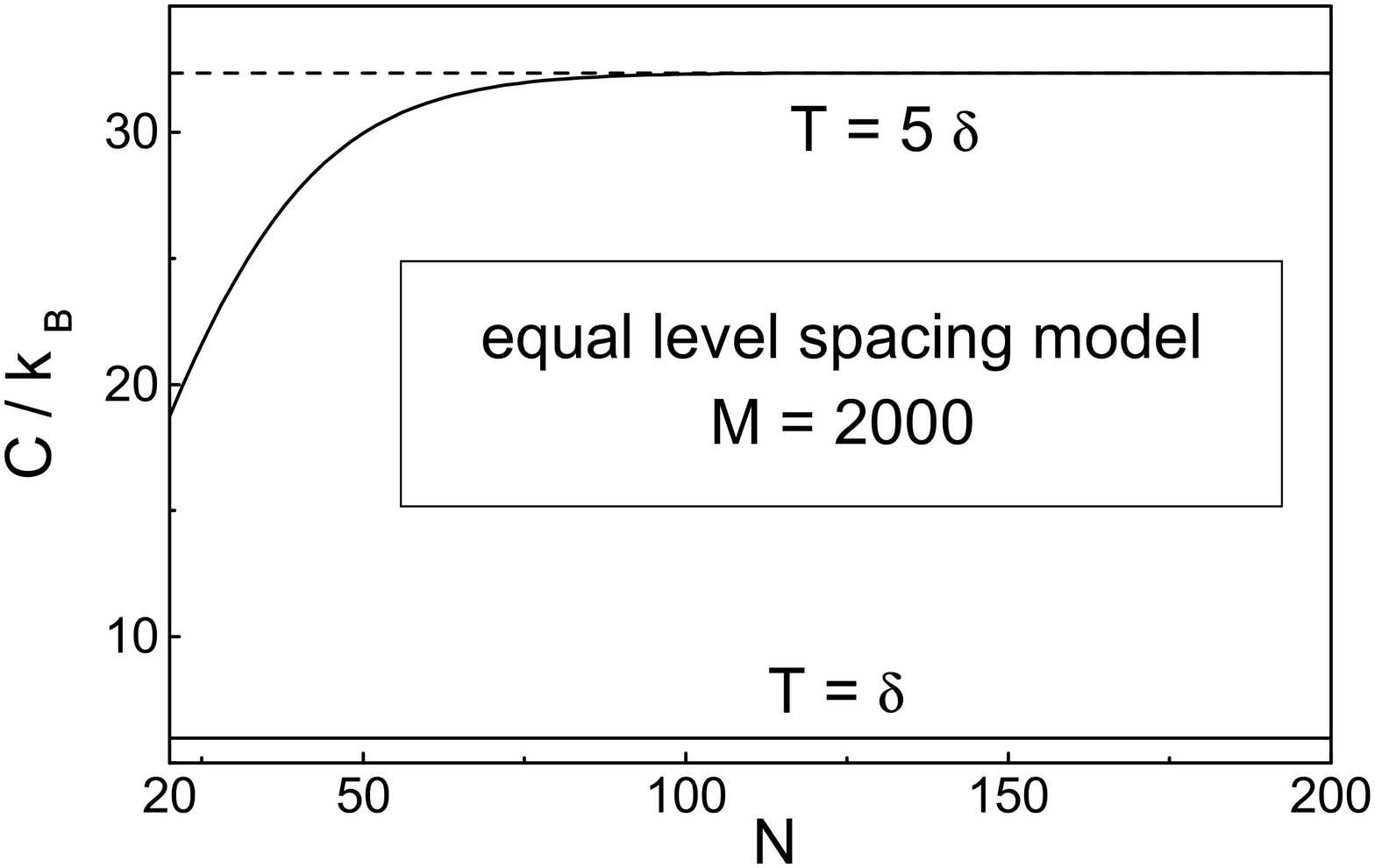}}
\caption{{\label{CvsNT5}} The canonical electronic heat capacity
vs the particle number in the equal level spacing model ($\delta$
is the level spacing). Solid lines represent the results obtained
by our method. The dashed line represents the result of
Ref.~\cite{denton}. At $T=\delta$ the exact canonical results
coincide with the results of Ref.~\cite{denton}.}
\end{figure}

As shown in Appendix the exact canonical partition function for
the equal level spacing model turns out to be much more
complicated than the expression given in Ref.~\cite{denton} in
approximation $M>N\gg 1$. However the difference reveals itself
mainly at small particle number and $T/\Delta >1$. This is
demonstrated in Figs.~(\ref{mdepend}),(\ref{CvsNT5}) where
$C_{can}$ is displayed v.s. $N$ and $T$.


\section{Conclusion}

We have shown that the canonical partition function $Z_N$ which in
general case is a symmetric polynomial of the $N$-th order (in
variables of $\exp\{\beta(\lambda-\varepsilon_s)\}$) can be
expanded in polynomials the orders of which range from $0$ to $N$.
Such representation of $Z_N$ takes into account that in the
independent particle model the full configuration space is
naturally divided by the Fermi surface and each excitation in this
model is generated by particle-hole pairs the amount of which
range from 0 to $N$ as well. Therefore the polynomial expansion of
$Z_N$ is a sum over products of particle and hole symmetric
polynomials the order of which vary in the same manner as the
amount of pairs in the excitation.

In consequence the canonical averages such as the occupation
numbers, heat capacity and magnetic susceptibility are expressed
via similar sums involving the same polynomials and others defined
in narrower spaces (where a part of states is omitted). We have
suggested recurrent relations reduced the calculations of $n$-th
order particle (hole) polynomials, corresponding to distributions
of $n$ particles (holes) over states above (below) the Fermi
level, to first order polynomials.

Though the sum  $Z_N$ stretches to $N$-th order polynomials, its
convergence is so high that this version of the exact canonical
approach can be employed for mesoscopic systems in a wide range of
temperatures and particle numbers. Moreover , as only few first
terms bring in a real contribution to canonical averages in the
temperature range of several level spacings near the Fermi level,
the elaborated method is especially effective for studying
structure and size effects in the temperature regime which is most
appropriate to this goal.

\appendix
\section{The canonical partition function of the equal
spacing model.}

In this model each single electron level in the applied magnetic
field $B$ gains the Zeeman splitting
 \begin{equation}
 \varepsilon_s^{\pm}=s\Delta\pm\omega\label{app1}
 \end{equation}                         
 $\Delta$ is the level spacing, $s=0,1,\ldots ,M-1$, where $M$
 is the amount of the electron levels, $M>N$. $N$ is the electron
 number. Thus each polynomial in Eq.~(\ref{Eq35} is defined in the
 space of $q_s$ (here we set $\lambda=0$)
 \begin{equation}\label{app2}
 q_s=q^s, \;\;\;
 q=exp(-\beta\Delta),
 \end{equation}                         
 that allows a polynomial $[[k]]$ to be calculated analytically
 \begin{eqnarray}
 [[k]]  & = & q^{k(k-1)/2}Q(M,k),\label{app3} \\
 Q(M,k) & = & \delta_{k.0}+(1-\delta_{k.0})
 \prod_{\kappa=1}^{k}(1-q^{M+1-\kappa)}/(1-q^{\kappa})
 \nonumber
 \end{eqnarray}                         
Eq.~(\ref{app3}) is straightforwardly found for all $k$ at small
$M$ and then extended for any $M$ by induction. Substituting
Eq.~(\ref{app3}) into Eq.~(\ref{Eq35}), where $n$ has to be
replaced by $N$, one obtains
 \begin{equation}
 Z_N = \left[ N\right]=exp\{-\beta E_0(N)\}\widetilde{Z_N},
 \label{app4}
 \end{equation}                       
 \begin{eqnarray*}
 \widetilde{Z_N} =
 \sum_{\nu=0}^{\frac{[N-\xi_N]}{2}}
 \left \{ 2-[1-\xi_N]\delta_{\nu.0} \right \}q^{\nu(\nu+\xi_N)} \\
 \times Q(M,\frac{N+\xi_N}{2}+\nu)
 Q(M,\frac{N-\xi_N}{2}-\nu)
 \cosh\{ \beta(2\nu+\xi_N)\},\\
\xi_N=\frac{1-(-1)^n}{2}.
 \end{eqnarray*}

 $E_0(N)$ is the ground state energy in this model at $T=B=0$;
 $E_0(N)=\Delta\left [ N(N-2)+\xi_N \right ]/4$.

 Eq.~(\ref{app4}) is the exact partition function. In order to crossover
 to the expression given in Ref.\cite{denton} two
 approximations have to be done. Firstly the amount of the levels $M$ has
 to be infinitely large, $M\rightarrow\infty$, then
 \begin{displaymath}
 Q(M,k)\rightarrow Q(\infty ,k)=\delta_{k.0}+
 (1-\delta_{k.0})\prod_{\kappa=1}^{k}(1-q^{\kappa})^{-1}.
 \end{displaymath}
 Secondly the values of $q=exp(-\beta\Delta)$ must be small enough
 whereas the particle number $N\gg 1$ then the dependence $Q$ on
 $\nu$ in Eq.~(\ref{app4}) can be dropped
 \begin{displaymath}
 Q(\infty ,\frac{N\pm\xi_N}{2}\pm\nu)\rightarrow
 \prod_{\kappa=1}^{\infty}(1-q^{\kappa})^{-1}.
 \end{displaymath}

 \end{document}